\documentstyle[sprocl]{article}

\def\ra{\rightarrow}

\begin{document}
\title{The Saga of $h_c$ and $h_b$ Search in Heavy Quarkonia}

\author{T. Barnes}
\address{Physics Division, Oak Ridge National Laboratory\\
      Oak Ridge, TN 37831-6373 \\ and
               Department of Physics \\
           University of Tennessee  \\ Knoxville, TN 37996-1501}

\author{T.E. Browder and S.F. Tuan}
\address{Department of Physics \\
 University of Hawaii at Manoa \\
              Honolulu, HI 96822}

\maketitle
\begin{abstract}

In this brief review, we examine carefully and critically the status of search
for the $h_c$ [$\psi(1^1P_1)$] charmonium state and the $h_b$
[$\Upsilon(1^1P_1)$] bottonium state, initiated by the E760 experiment.
Recent experimental studies at CLEO/BABAR/BELLE are examined in the light of
new theoretical and phenomenological understanding.
\\~\\

PACS numbers(s):  13.20.Gd, 13.75.Cs, 14.40.Gx
     \end{abstract}

\section{\bf Introduction}
Using a circulating monochromatic antiproton beam at the ISR
and a hydrogen gas jet target,
experiment R704 studied $p\bar{p}$ interactions at the center of mass
energy range of interest for charmonium physics, with unprecedented energy
resolution. However, before significant data were taken the ISR was closed down
and dismantled. Experiment E760 was resurrected deploying the cooled antiproton
beam at Fermilab using much of the same technique
as the original CERN experiment.
Here much data were accumulated of relevance to charmonium physics in a
variety of areas \cite{E760a,E760b,E760c}. We examine in this brief review the
E760 result on $h_c[\psi(1^1P_1)]$, the sequel E835 follow up, and recent
experimental studies at CLEO/BABAR/BELLE in the light of new theoretical and
phenomenological understanding.

\section{\bf{E760 results on the charmonium $\psi(^1P_1)$}}

The E760 measurements of $\psi(^1P_1)$ are summarized
in Table~\ref{measurements} \cite{E760c}.
These results have attracted
theoretical attention \cite{Murgia} with a calculation
of the decay width for the $\psi(^1P_1) \ra p \bar{p}$ process
considering only the constituent quark mass correction,
concluding that this kind of correction leads to
$\Gamma(^1P_1 \ra p \bar{p})$ in the range 1 - 10 eV.
This is substantially smaller
than an earlier conservative estimate \cite{Kuang1} which sensibly
normalized the
$^1P_1 \ra p \bar{p}$ estimate to that of $\eta_c \ra p \bar{p}$
decay since both processes
violate the helicity selection rule \cite{Brodsky} but data is available
 on the latter
process, obtaining 186 eV. We note however that in \cite{Murgia} Murgia's
final expression for $\Gamma(^1P_1 \ra p \bar{p})$ [c.f. his Eq. (34)],
based on constituent quark
mass corrections to the usual massless QCD models for exclusive processes,
has a multiplicative
dependence on $[M(\chi_{c2})/M(\psi(^1P_1)]^{12}$ and hence is
very sensitive to (not very well known) mass values for these charmonium
states. In E760 \cite{E760c} there was the incorrect attribution
that Kuang-Tuan-Yan \cite{Kuang1}
predicted $\Gamma(^1P_1 \ra J/\psi + \pi^o) \sim 2$ keV.
The actual phenomenological approach yields for this rate
the value $0.3(\alpha_M/$
$\alpha_E)$ keV (c.f. Eq. (35) of \cite{Kuang1})
where $\alpha_M$ and $\alpha_E$ are the magnetic
and electric couplings in the multipole expansion approach. To maintain
theoretical reasonableness (absence of large anomalous magnetic moment for
quarks \cite{Dalitz}) and the experimental
constraint on $\Upsilon (3S) \ra \Upsilon(^1P_1)\pi\pi$
from CLEO \cite{Tuan1},
we must set $\alpha_M = \alpha_E$. Thus $\Gamma (\psi(^1P_1) \ra
J/\psi+\pi^o)$ = 0.3 keV. This is reassuring since it reduces this
isospin violating decay which seems too large in E760 \cite{E760c}.
With this value, we infer in the language of E760
\cite{E760c} that $BR(^1P_1 \ra p \bar{p}) \sim 4.33 \times 10^{-4}$ for a
total $\Gamma(^1P_1)$ width of say
700 keV suggested by E760 \cite{E760c}, which is then order of magnitude wise
quite consistent with $\Gamma (\psi(^1P_1) \ra p \bar{p})$ = 186 eV
estimated by \cite{Kuang1},
but at variance with the recent calculation of Murgia \cite{Murgia}.

     We continue to find the E760 result \cite{E760c}
\begin{equation}
BR(\psi(^1P_1) \ra J/\psi+\pi^o)/BR(\psi(^1P_1)
\ra J/\psi+\pi\pi) > 5.5 (90\% C.L.)
\end{equation}
problematic.  Isgur et al. \cite{Isgur1} estimated that
$BR(\psi(^1P_1) \ra J/\psi+\pi^o) \sim 10^{-3}$ to $10^{-4}$,
while Bodwin, Braaten, and Lepage \cite{Bodwin} pointed out
that the rate for $\psi(^1P_1) \ra J/\psi+\pi\pi$ has been
estimated within a well-developed phenomenological framework 
\cite{Kuang1} to be of order of 6 keV,
leading to $BR(\psi(^1P_1) \ra J/\psi+\pi\pi)  \sim 10^{-2}$. One of us 
\cite{Tuan2} noted that E760 \cite{E760c} also did
not observe  $\psi(^1P_1) \ra \gamma \eta_c$, a
mode expected to be dominant
with a branching fraction $\sim$ 50\% by any reasonable
estimate.  This could be due to the
small $BR(\eta_c \ra 2 \gamma)  \sim 3 \times 10^{-4}$
branching fraction which strains
the limits of E760 capacity to detect final state
photons from $\eta_c( \ra \gamma \gamma) \gamma$. It remains strange that
$\psi(^1P_1)$ is discovered via this peculiar isospin violating mode involving
single pion emission, while
the isospin allowed dipion mode and the dominant $\gamma \eta_c$
mode remain to be
identified. We recognize that a theory based on
the stress-energy tensor in QCD can lead to
dominance of single pion over dipion emission \cite{Voloshin}.

The $\psi(^1P_1)$ was confirmed via the same mode in
another hadron initiated experiment, E705, \cite{Antoniazzi} at the same
mass. However, for increasing the signal-to background ratio, they
imposed the cut $M_{\pi \pi} > 80\%$ on the dipion mass
distribution inspired by the $M_{\pi \pi}$ distribution in the channel
$\psi' \ra J/\psi \pi \pi$.  Kuang \cite{Kuang2} pointed out
that this cut is not suitable for $\psi(^1P_1) \ra J/\psi \pi \pi$
since the $M_{\pi \pi}$ distribution for this process is not
analogous to that in $\psi' \ra J/\psi \pi \pi$ but is analogous
to the KTY \cite{Kuang1} dipion mass distribution for $\Upsilon (3S) \ra
\Upsilon (^1P_1) \pi \pi$ which is strongly peaked at the low dipion
mass region.  Therefore the E705 experimental cut
actually eliminates 73\% of the signal events, so that it eliminates the
chance of seeing this dipion decay mode of $\psi (^1P_1)$.

     The position for the $\psi(^1P_1)$ charmonium state
is expected to be close to the center of gravity (c.o.g)
of the $^3P_J$ states
if the spin-spin contribution arising from the Fermi hyperfine interaction
(due to one gluon exchange between $c$ and $\bar{c})$ can be
neglected \cite{Dalitz}. Halzen et al.~\cite{Halzen}
performed a one-loop perturbative correction to the
$^1 P_1$-$^3 P_{c.o.g.}$ and found it to be $+0.7\pm 0.2 \ MeV$
in accord with E760 \cite{E760c}, thus
here the experiment does follow the conventional theoretical expectation.

However, Isgur \cite{Isgur2} is surprised by the smallness of the observed
splitting, because it is difficult
to understand why the non-perturbative couplings of the P-waves to virtual
decay channels would not produce relative shifts amongst these states
of order 10 MeV. Isgur notes that it
would be interesting to understand
whether the decoupling from these virtual channels is really as complete as
it would appear from the Halzen et al. \cite{Halzen} result (even though the
$D^0 \bar{D}^{*0}$ continuum at 3871.2 MeV is only a few hundred MeV
away) or if this experimental/theoretical
degeneracy \cite{E760c,Halzen} is mainly accidental. We note that Isgur's
conjecture \cite{Isgur2} of an up to 10 MeV (upward) shift
in the position of $\psi(^1P_1)$ is supported by the
hyperfine spin-spin correction calculation of Chen and Oakes
 \cite{Chen} who find E$(^1P_1)$ - E$(^3P_{c.o.g.})$ = 4 to 6 MeV
which is several times larger than the E760 experimental value \cite{E760c} of
$0.93 \pm 0.28$ MeV. This also suggests that the
agreement of the  one-loop perturbative
calculation \cite{Halzen} with the E760 data is probably fortuitous. It is
well to remember that for $\psi(^1P_1)$ with an estimated width $<$ 1.1 MeV,
E760 \cite{E760c} only scanned
[in increments (decrements) of 0.5 MeV] around the c.o.g.
3525 MeV of $^3P_J$ by about 12 MeV (i.e. $3525 \pm  6 MeV)$, hence a
mass shift of 10 to 6 MeV \cite{Isgur2,Chen} from c.o.g. would
have escaped their search!
[Note E705 \cite{Antoniazzi} has a 8  MeV$/c^2$ mass uncertainty
in their observation of
$^1P_1 \ra J/\psi+\pi^o$ which reflects the statistical
uncertainty in measuring c.o.g. of $\chi_{cJ}$ states.
At $\pm 8 \ MeV$ E705 is already outside
range explored by E760.]  In Table~\ref{measurements} E760 measurements
on $\psi(^1P_1)$ are compared with theory.
\begin{table} [ht]
\caption{E760 measurements on $\psi (^1P_1)$ are compared with
theoretical estimates.}  
\label{measurements}

\begin{tabular}{ll} \\ \hline
E760 measurement of \cite{E760c}&

Theoretical estimates \cite{Isgur1,Bodwin,Kuang1} \\
$\Gamma(\psi (^1P_1) \ra J/\psi + \pi^0) / 
\Gamma(\psi (^1P_1) \ra J/\psi
+ \pi \pi)$    & 1/10 to 1/100 \\
&  
an earlier \cite{Voloshin} {\underline{much}} larger  \\
$>$  5.5 [90\% C.L.]                 & estimate discussed in text  \\ \hline
E760 measurement \cite{E760c}	  &  Theoretical~estimate \cite{Isgur2,Chen}  \\
$\Delta M = M(\psi(^1P_1)) -M_{c.o.g.} (\chi_{cJ})$  &   $\geq$ 4 to 10
MeV \\ & One loop perturbation \cite{Halzen}  \\
0.93 $\pm 0.28$ MeV& 0.7 $\pm$ 0.2 MeV  \\ \hline
\end{tabular}
\end{table}

It has been noted \cite{Barnes2} that vector confinement would be
strongly ruled out by
the discovery of the $^1P_1 \ h_c$ spin-singlet $c \bar{c}$ charmonium
state by the E760 Collaboration \cite{E760c}; with
typical $c \ \bar{c}$ potential
model parameters, the vector confinement spin-spin term
predicts a splitting between the c.o.g. of
the $(\chi_{cJ}) \ ^3P_J$ triplet states and
the $h_c \ ^1P_1$ of order 30 MeV,
whereas E760 \cite{E760c} finds that the splitting is about 1 MeV. Scalar
confinement on the other hand would give
a very small $h_c - ^3P_{c.o.g.}$ splitting, because
there is no scalar spin-spin interaction,
and the non relativistic OGE spin-spin interaction is a contact term.
Indeed the {\bf{L.S.}} pattern of the $\chi$ states looks like
scalar confinement more than vector, independent of the location of the $h_c$.
So the proposed 6-10 MeV shift \cite{Isgur2,Chen} of $h_c$ from
$^3P_{c.o.g.}$ would raise the very interesting issues of both modification of
the naive $c\bar{c}$ spectrum and
the relative proportion of vector vs. scalar components of the confinement
potential. We remind that the shift is upward from the c.o.g. value because
of the (non relativistic) Stubbe and Martin theorem \cite{Stubbe}.

Since 1998 there has been a recent paper by Qiao and Yuan \cite{Qiao} which
states that the statistical significance of the E760 $\psi(^1P_1)$ \cite{E760c}
is only a slightly more than $3 \sigma$ signal with no other experiments
confirming it (the E705 \cite{Antoniazzi} confirmation is now in doubt as
discussed
in this section).  Hence the existence of  $\psi(^1P_1)$ is still based
on very weak experimental signals.  The authors \cite{Qiao} proposed innovative
methods of finding  $\psi(^1P_1)$ at HERA-B.  However a search strategy
at HERA need not be based on E760's  $\psi(^1P_1) \rightarrow J/\psi + \pi^0$
``discovery'' mode, as discussed in the present section.  Finally there is the
E835 upgrade of the E760 experiment. The design of the E835 detector is
basically the same \cite{Peoples} detector as E760 (emphasizing photon
detection), but with improvements in tracking and data acquisition capability.
With significantly higher statistics the succeeding E835 experiment was unable
to confirm E760's enhancement in 
$p+\bar{p} \ra J/\psi+\pi^0$ at $\sqrt{s}\ =
3526.2$ MeV which was supposed to be a candidate for $h_c(^1P_1)$
 \cite{Patrignani}.

\section{\bf Study of $h_c$ at BELLE/BABAR}

A promising approach for the detection of the $h_c$ has recently been proposed
by Suzuki and Gu \cite{Suzuki}. Suzuki suggests looking for the $h_c$ by
measuring the final state $\gamma\eta_{c}$ of the cascade $B \ra h_{c}K/K^{*}
\ra \gamma\eta_{c}K/K^{*}$. This channel is especially timely given the
announcement by the BELLE Collaboration of the $\eta_{c}(2S)$ in B-decays
 \cite{Choi} and, previously, the observation of the related decay, $B \ra
\chi_{c0}K$. \cite{Abe} That the factorization-forbidden decay $B \ra
\chi_{c0}K$ occurs as vigorously as the factorization-allowed decays to other
charmonia e.g. $B \ra \chi_{c1}K$. \cite{Abe} On the basis of this finding, we
expect that another factorization-forbidden decay $B \ra h_{c}K$ may occur
just as abundantly as $B \ra \chi_{c0}K$. Since $h_c \ra \gamma\eta_c$ is one
of the two main decay modes of $h_c$, the decay 
$B \ra h_{c}K$ cascades down to
the final state $\gamma\eta_{c}K$ about half the time. The only background
for this process at the B factories will be the process 
$B \ra \psi^{\prime}K
\ra \gamma\eta_{c}K$. Since the branching fraction for $\psi^{\prime} \ra
\gamma\eta_c$ is miniscule, this background is two orders of magnitude smaller
than the signal. If one can reconstruct $\eta_c$ from $K\bar{K}\pi$ or by
$\eta\pi\pi$ with say 50\% efficiency, $10^7$ B's translate to roughly 100
events of the signal. Therefore in principle, we have a very good chance to
observe $h_c$ through $B \ra \gamma\eta_{c}K$. For $B(B \ra h_{c}K) \sim
B(B \ra \chi_{0}K$, Suzuki \cite{Suzuki} estimated the cascade branching
fraction
\begin{equation}
B(B^+ \ra h_c K^+ \ra \gamma \eta_c K^+ \ra \gamma (K \bar{K} \pi) K^+)
\sim 2 \times 10^{-5}
\end{equation}
Gu's estimate \cite{Suzuki} of the cascade branching fraction is
\begin{eqnarray}
B( B^+ \ra h_c K^+ \ra \gamma \eta_c K^+ \nonumber \\ 
\ra \gamma ( K^0_S K ^+ \pi^- + c.c. ) K^+ \nonumber \\
\ra \gamma ( \pi^+ \pi^- K^+ \pi^- + c.c. ) K^+ ) \simeq 
3.5 \times 10^{-6}
\end{eqnarray}
With $10^8$ B's, and an efficiency $\epsilon = 10\%$ \cite{Aubert}, there will
be about 35 events of the $h_c$.

The largest (and the only) uncertainty \cite{Suzuki1} appears to be in the
$B \ra h_{c}K/K^{*}$ branching ratio. The heuristic assumption of a sizable
branching for $B \ra h_{c}K/K^{*}$ was based on the experimental discovery of
$B \ra \chi_{c0}K$ by BELLE/BABAR. \cite{Abe} The main difference between
$\chi_{cJ}$ (J=0,1,2) and $h_c$ is not spins but charge parity. C=+ for
$\chi_{cJ}$ seem copiously produced, but $C=-$ for $h_c$ is not yet seen. It is
difficult to know how this difference in C plays out in the dynamics of decay,
and theorists have no reliable way to compute it (if they are careful and
honest). It is nevertheless of interest that theoretical work \cite{Fazio}
obtain for the $B^{-}$ chain of Eq. (2) a value $(4-26)\times 10^{-6}$ where
the upper end is actually consistent with Suzuki's estimate. As stated in
Suzuki \cite{Suzuki}, one will have to work harder for precision of the $h_c$
mass, meaning the resolution of the $h_c$ mass. Just identifying $h_c$ at the
B factories should not be difficult if it is there. We note also the
phenomenological work of Eichten {\it et al.} \cite{Eichten} on $B \ra Kh_c
\ra K\gamma(500 \ MeV)\eta_c$ that they estimate 11.7 K events for this chain.
For comparison they noted that the sample that yielded the $39 \pm 11$
$\eta_{c}^{\prime}$ discovery events by BELLE \cite{Abe} was (encouragingly)
about 30 K events.

Eichten {\it et al.} \cite{Eichten} estimate that for inclusive $B(B \ra
h_{c}X) \sim \ 0.132 \pm 0.06 \%$, while Beneke {\it et al.} \cite{Beneke}
predicted that for inclusive $B(B^- \ra h_{c}X) \sim 0.13 - 0.34 \%$ with
production of $c\bar{c}$ pair in the color octet state. However it was
pointed out by De Fazio \cite{Fazio} that with their estimate of $B(B^-
\ra K^{-}h_{c}) = (2 - 12) \times 10^{-4} $ and Suzuki's estimate given by
Eq. (2), the exclusive mode represents already a sizeable fraction of the
inclusive $B^- \ra Xh_c $ decay. Hence it will make sense to look for the
inclusive $B \ra h_c X$ production at BELLE/BABAR if exclusive $B \ra
h_{c}K/K^{*}$ is significantly lower than the Suzuki \cite{Suzuki} and
De~Fazio~\cite{Fazio} estimates.

It has been pointed out \cite{Swanson} that the radiative width for
$\eta_{c2}(1 ^1D_2) \ra h_{c}\gamma$ is about 340 keV, while that for
$h_c \ra \eta_{c}\gamma$ is about 350 - 500 keV. There are {\bf no radial
nodes}, so these should be fairly reliable predictions. So, it looks like
both $E1$ modes will be LARGE branching fractions, since the total widths
for $\eta_{c2}$ and $h_c$ are expected to be of 1 MeV scale. It would be a
funny way to find the $h_c$, once we find the $\eta_{c2}$ experimentally!

There is a potential candidate for $\eta_{c2}$ at $X(3872)$, \cite{Choi1}
which is indeed not far from the nonrelativistic $c\bar{c}$ potential
model prediction for $\eta_{c2}(^1D_2)$ at 3799 MeV. \cite{Barnes3} An useful
test\cite{Choi2} for $X = \eta_{c2}(2^{-+})$ is that the $\eta_{c2} \ra
\pi^+ \pi^- \eta_{c}$ and $\gamma h_{c}$ decays are allowed and expected to
have widths in the range of 100's of keV,\cite{Godfrey} and much larger
than that for the isospin violating $\pi^+ \pi^- J/\psi$ mode. If the $X(3872)$
were $\eta_{c2}$, the total exclusive branching fraction for $B^+ \ra
K^+ \eta_{c2}$ decay, which is non-factorizable and suppressed by an $L=2$
barrier, would be anomalously large, typically\cite{Olsen} $B(B^+ \ra
K^+\eta_{c2})$ $\gg 1.3 \times 10^{-3}$. Direct search for $X(3982) \ra
\pi^+ \pi^- \eta_c $ is currently underway at BELLE.\cite{Olsen} From the
viewpoint of $h_c$ search, this surpasses a search for $X(3872) \ra
\pi^0 \pi^0 J/\psi$,\cite{Godfrey},\cite{Voloshin1} where absence of this
mode would imply C=(+) with $c\bar{c}$ candidates in 
$\eta_{c}^{\prime\prime}$, $\chi_{c1}^{\prime}$ as well as $\eta_{c2}$
(it could be consistent also with the deuson model \cite{Tornqvist}
with $J^{PC} = 0^{-+}, 1^{++}$ which remains viable particularly as $X(3872)$
sits on the $D\bar{D^*}$ threshold). Steve~Olsen~\cite{Olsen} has pointed out
that $X(3872) \ra \pi^0 \pi^0 J/\psi$ is damned hard experimentally! Moreover,
the signal, if it is there, is only half of $\pi^+ \pi^- J/\psi$; the $\pi^0$
efficiency is smaller than that for charged $\pi$ and the resolution is
much much worse. Hence it will be quite a while before any useful result from
$\pi^0 \pi^0 J/\psi$ emerges from BELLE/BABAR. Again we note that Eichten
{\it et al.} \cite{Eichten} say that $B \ra K\eta_{c2} \ra
K\gamma(280 MeV) h_c \ra K\gamma(280 MeV)\gamma(500 MeV) \eta_c $,
8.1 K events arise (remember again that the sample that yielded the
$39 \pm 11$ $\eta_{c}^{\prime}$ discovery events~\cite{Choi} was about 30
K events), hence they hold out for simultaneous observation of $\eta_{c2}$
and $h_c$. We note that the BABAR collaboration \cite{Choi1} in setting an
upper limit for the narrow ($\Gamma \leq 1 MeV) \ h_c$ state in the decay
$B^- \ra h_{c}K^- $, and $h_c \ra J/\psi \pi^+ \pi^- $, did not deploy $h_c
\ra \gamma\eta_c $ (with branching ratio $ \geq 50\%$), whereas
$h_c \ra J/\psi \pi^+ \pi^- $ including $ m_{\pi} \neq 0$ corrections
discussed by one of us \cite{Tuan1} is estimated to have miniscule partial
width $\Gamma(h_c \ra J/\psi \pi^+ \pi^-) \sim 1.07$ keV. However, we believe
the BABAR collaboration \cite{Choi1} is primarily interested in $X(3872) \ra
J/\psi \pi^+ \pi^- $ decay measurements.

Since the predicted radiative partial width for $h_c \ra \gamma\eta_c $ is
especially large, which suggests that decays to $\gamma\eta_c $ may provide
a discovery channel for the elusive $h_c$. One possibility \cite{Barnes3} is
$\gamma\gamma \ra \eta_{c}^{\prime}$, followed by the decay chain
$\eta_{c}^{\prime} \ra \gamma h_c $, $h_c \ra \gamma\eta_c $. Since these
electromagnetic couplings are all reasonably well understood, detection of
$h_c$ would simply be a matter of accumulating adequate statistics at a high-
energy $e^{+}e^{-}$ facility (e.g. BELLE/BABAR) or even at RHIC, with
sufficient $\gamma\gamma$ luminosity at $\sqrt{s}$ = 3.7 GeV. Indeed
CLEO, \cite{Asner} before being reconstructed to become a lower energy
$e^{+}e^{-}$ facility CLEO-c, had established $\gamma\gamma \ra
\eta_{c}^{\prime}$ and using the method of Barnes {\it et al.}, \cite{Browder}
concluded that $\Gamma_{\gamma\gamma}(\eta_{c}^{\prime}) = 1.3 \pm 0.6$ keV.
This is at most a factor of 3 lower than quark model estimate \cite{Ackleh}
with $\Gamma_{\gamma\gamma}(\eta_{c}^{\prime}) = 3.7$ keV.
Munz \cite{Munz} summarized/obtained other predictions between a
factor of 2 to 3 smaller than the quark model estimate.~\cite{Ackleh}

\section {\bf Study of $h_b$ at CLEO}

The data set consists of $5.8 \times 10^6 \ \Upsilon(3S)$ decays observed
with the CLEO III detector at the Cornell Electron Storage Ring
(CESR). \cite{Bonvicini} Hence the CLEO sample of hadronic $\Upsilon(3S)$
decays ($\sim 3.5 \times 10^6$) will be nearly 15 times \cite{Rosner} their
previous total some ten years ago. The most straightforward search for $h_b$
was proposed \cite{Kuang1} as $\Upsilon(3S) \ra \pi^+ \pi^- h_b $, where the
mass of $h_b \geq 9900 \pm 0.17$ MeV c.o.g. value of $^3P_J$ mass values
according to the Stubbe-Martin theorem. \cite{Stubbe} The standard multipole
model \cite{Kuang1} predicts for $\Upsilon(3S) \ra \pi^+ \pi^- h_b$ a spectrum
in which the $\pi\pi$ mass distribution is strongly peaked at the lower
$M_{\pi\pi}$ end. As pointed out by Rosner \cite{Rosner1} this comes from
the fact that the $\pi\pi$ system should be $J^P = 0^+ $, and the
$1^- \ra 0^+ 1^+ $ transition must proceed with final state L=1 between the
dipion and the $h_b$ state. The peaking peculiar to this transition, may allow
one to {\bf strengthen the $h_b$ signal by selecting dipions with particularly
low mass}. In the case of the earlier CLEO work of F. Butler {\it et al.}
, \cite{Tuan1}, an upper limit of $B(\Upsilon(3S) \ra \pi^+\pi^- h_b < 0.18\%$
at $90\%$ confidence level was established. The standard multipole model
estimate given by one of us \cite{Tuan1} is that $B(\Upsilon(3S) \ra
\pi^+ \pi^- h_b)$ should bracket the range $0.022\%$ to $0.08\%$, while
Voloshin \cite{Voloshin} will predict a further factor of 0.05 reduction over
th standard multipole model value. With nearly 15 times more hadronic
$\Upsilon(3S)$ accumulated since ten years ago, it will be interesting to
push the $B(\Upsilon(3S) \ra \pi^+ \pi^- h_b)$ limit further, but deploying
the Rosner \cite{Rosner1} suggestion of selecting dipions with particularly
low mass, to enhance a possible $h_b$ signal. A {\it caveat} is appropriate when
dealing with radial excitation states like $\Upsilon(3S)$ and
$\psi(2S)(\psi^{\prime})$. Rosner \cite{Rosner2} pointed out that suppression
of $\rho\pi$, $K^{*}\bar{K}$ decays in the famous $\rho-\pi$ puzzle, could be
due to a radial node in $\psi(2S)$. Could there be a similar suppression in
say $\Upsilon(3S) \ra h_b \pi^+ \pi^- $ or $\psi(2S) \ra h_c \pi^0 $ recently
proposed by Kuang \cite{Kuang3} as a search method for $h_c?$ Of course
$\psi(2S) \ra J/\psi \pi\pi $ and $\Upsilon(3S) \ra \Upsilon(1S) \pi\pi $ do
exist, but are they to be regarded as large or small in some context, let
alone that $h_c$ and $h_b$ represent different exclusive 
channels from radially
excited $\psi$ and $\Upsilon$ to $J/\psi$ and $\Upsilon(1S)$ respectively?
The calculation of Ackleh and Barnes \cite{Ackleh} suggests that radially
excited $\eta_{c}(2S) \ra \gamma\gamma $ is at most a factor of three higher
than experimental measurement, \cite{Asner} so here radial node suppression
is relatively modest. With respect to $\psi^{\prime} \ra h_c \pi^0 $ there
could be lack of phase space (not always a problem though since $\psi(4040)
\ra D^{*}\bar{D^{*}}$ is dominant with essentially no phase
space \cite{Barnes3}) and isospin violation as well (not a problem for
Voloshin \cite{Voloshin} though). However at BES II, the accumulation of
$\psi(2S)$ is only about $1.4 \times 10^7 $ events, \cite{Harris} hence
Kuang's estimate \cite{Kuang3} of $\psi^{\prime} \ra h_c \pi^0 $ based on
$3 \times 10^7 \psi^{\prime}$ (and an efficiency $\epsilon = 10\% $) needed
to be reduced accordingly.

There could be an indirect test of the validity of the standard multipole
Kuang-Yan model \cite{Yan} outside of $h_c$ and $h_b$, in terms of their prediction
that $\Upsilon(1D) \ra \Upsilon(1S) \pi \pi $ is about 24 keV. Here the
CLEO experiment \cite{Bonvicini} obtained for
$B(\Upsilon(3S) \ra \gamma\gamma\Upsilon(1D))B(\Upsilon(1D_{J}) \ra
\pi^+ \pi^- \Upsilon(1S)) < 2.7 \times 10^-4 $ for a sum over all different
$J_{1D}$ values. This upper limit is inconsistent (lower by a factor of about
7) with the rate estimated by Rosner \cite{Rosner3} using the Kuang-Yan model
for $\Gamma(\Upsilon(1D) \ra \pi^+ pi^- \Upsilon(1S))$, \cite{Yan} and a factor
of about 3 higher than the predicted rate based on the model by Ko. \cite{Ko}
The CLEO upper limit \cite{Bonvicini} is about 30 times higher than those
predicted by Moxhay's model \cite{Moxhay} which uses the Voloshin \cite{Voloshin}
approach. There is then the BES measurement \cite{BES} of the
$\psi^{\prime\prime} \ra J/\psi \pi \pi $ rate (supportive of
Kuang-Yan \cite{Yan}). Godfrey \cite{Godfrey1} pointed out that the reduced
rate for the $c\bar{b}$ system found by rescaling the BES
measurement, \cite{BES} is considerably larger than the rate found by rescaling
the $\Upsilon(1D) \ra \Upsilon(1S)\pi\pi $ CLEO limit. \cite{Bonvicini}
However Godfrey \cite{Godfrey1} still believes that it is likely one can
reconcile the $b\bar{b}$ and $c\bar{c}$ results by properly taking into
account $2 ^3S_1 - 1 ^3D_1 $ mixing.

It is believed that the $\Upsilon(1D)$ found by CLEO at
10,161 MeV \cite{Bonvicini} is likely the $\Upsilon(1 ^3D_2)$ since there is
good agreement with lattice QCD calculations and potential models. As stressed
by Rosner, \cite{Rosner} since the $B\bar{B}$ threshold is quite far from
$\Upsilon(1D)$, coupled channel effects should be small, so the potential
models should be reliable. The prediction is probably good to within 0.02 GeV,
with 10.16 GeV the central value. Hence we expect the $\Upsilon(1 ^1D_2)$ to
lie in this neighborhood. Production of $\Upsilon(1 ^1D_2)$ could again lead
to large E1 transitions $\Upsilon(1 ^1D_2) \ra \gamma h_b $ and $h_b \ra
\gamma\eta_b $, for $h_b$ and $\eta_b $ search. However production of
$\Upsilon(1 ^1D_2)$ from $\Upsilon(3S)$ is a spin-flip transition. The
C-parity allows \cite{Rosner} $\Upsilon(3S) \ra \Upsilon(1 ^1D_2) + \gamma$ but
it is a highly hindered M1 transition, \cite{Sebastian} and is estimated to
have a partial width of 0.04 eV!

\section {\bf Concluding Remarks}

The short term objective on the search for $h_c$ should clearly be concentrated
on pushing further the limits on $B \ra h_c K/K^{*} \ra \gamma\eta_{c}K/K^{*}$
of the Suzuki-Gu \cite{Suzuki} approach at BELLE/BABAR. However Eq. (2) and
Eq. (3) may represent only the optimistic end of branching ratio expectations.
Inclusive $B \ra h_c X$ could also be considered. {\bf If} $X(3872) =
\eta_{c2}(1 ^1D_2)$, then $X(3872) \ra h_c \gamma \ra
\eta_{c}\gamma\gamma$ \cite{Swanson} would be a short cut towards 
$h_c$ discovery,
because of the large E1 branching ratios of each leg. An intermediate term
objective would be to use the CLEO III $5.8 \times 10^6 \ \Upsilon(3S)$ to push
lower the limit on $\Upsilon(3S) \ra \pi^+ \pi^- h_b $ with $h_b \ra
\gamma\eta_b $, though we understand \cite{Rosner} there are background issues.
Development of methods to identify the $1 ^1D_2(2^{-+})$ states of charmonium
[if not X(3872)] and bottonium are a priority, since we wish to deploy the
$^1D_2 \ra h_{c,b}\gamma \ra \eta_{c,b}\gamma\gamma$ favorable search method.
A long term objective \cite{Barnes3} at BELLE/BABAR (perhaps at RHIC) is to
accumulate adequate statistics, with sufficient $\gamma\gamma$ luminosity at
$\sqrt{s} = 3.7 GeV$, for a search to be conducted on $\gamma\gamma \ra
\eta_{c}^{\prime}$, $\eta_{c}^{\prime} \ra \gamma h_c $ (with partial width
$\sim 49$ keV), and $h_c \ra \gamma\eta_c $ (with partial width
$\sim 494$ keV).

\section*{\bf Acknowledgments}

One of us (S.F.T.) would like to thank M. Suzuki, Y.P. Kuang, F.A. Harris,
M.B. Voloshin (for his reading of a preliminary draft of this work in '96),
J.L. Rosner, and S.L. Olsen for helpful discussions. This work was supported
in part by the US Department of energy under Grant DE-FG-02-04ER41291 at the
University of Hawaii at Manoa, and Grant DE-AC05-96OR2264 managed by Lockheed
Martin Energy Research Corp. at Oak Ridge National Laboratory.

\end{document}